\documentclass[11pt,a4paper,twoside,groupcitations]{article}
\usepackage[T1]{fontenc}
\usepackage[ansinew]{inputenc}
\usepackage[english]{babel}
\usepackage{amsfonts}
\usepackage{amsmath}
\usepackage{bm}
\usepackage{array,upgreek,accents}
\usepackage{amsthm}
\usepackage{amssymb}
\usepackage{graphicx}
\usepackage{braket}
\usepackage{eucal}
\usepackage{verbatim}
\usepackage[table]{xcolor}
\usepackage{caption}
\usepackage{cite}
\usepackage{textcomp}
\usepackage{float}
\usepackage{multicol}
\usepackage{tikz}
\usetikzlibrary{positioning,arrows}
\usetikzlibrary{decorations.pathmorphing}
\usetikzlibrary{decorations.markings}
\usetikzlibrary{calc,decorations.markings}
\usetikzlibrary{arrows,shapes}
\usetikzlibrary{matrix,arrows}

\usepackage[colorlinks,hyperindex,unicode]{hyperref}
\definecolor{green1}{RGB}{0,128,0} 
\hypersetup{hidelinks,backref=true,pagebackref=true,hyperindex=true,colorlinks=true,breaklinks=true,urlcolor= blue}
\hypersetup{%
  colorlinks = true,
  linkcolor  = blue,
  citecolor = cyan,
}

\newcommand{\beq}{\begin{eqnarray}}
\newcommand{\eeq}{\end{eqnarray}}
\newcommand{\be}{\begin{eqnarray}}
\newcommand{\ee}{\end{eqnarray}}

\newcommand{\expec}[1]{\mbox{$\langle\, #1\,\rangle$}}

\renewcommand{\a}{\hat a}
\newcommand{\ac}{\hat a^{\dagger}}

\renewcommand{\d}{\mbox{${\rm d}$}} 
\newcommand{\lp}{\ell_{\rm p}}
\newcommand{\mpl}{m_{\rm p}}
\newcommand{\gn}{G_{\rm N}}

\newcommand{\Rh}{R_{\rm H}}

\newcommand{\Ng}{{N_{\rm G}}}

\newcommand{\Rinf}{{R_{\infty}}}
\newcommand{\dd }{{\mathrm d}}
\title{\bf Thermodynamic and configurational entropy of quantum Schwarzschild geometries}
\author{R.~Casadio$^{ab}$\thanks{E-mail: casadio@bo.infn.it},
$\ $
R.~da~Rocha$^{c}$\thanks{E-mail: roldao.rocha@ufabc.edu.br},
$\ $
A.~Giusti$^{d}$\thanks{E-mail: agiusti@phys.ethz.ch}
$\ $
and
P.~Meert$^{e}$\thanks{E-mail: pedro.meert@unesp.br}
\\
\\
$^a${\em Dipartimento di Fisica e Astronomia, Universit\`a di Bologna}
\\
{\em via Irnerio~46, 40126 Bologna, Italy}
\\
\\
$^b${\em I.N.F.N., Sezione di Bologna, I.S.~FLAG}
\\
{\em viale B.~Pichat~6/2, 40127 Bologna, Italy}
\\
\\
$^c${\em Federal University of ABC, Center of Mathematics}
\\
{\em Santo Andr\'e, 09210-580, Brazil.}
\\
\\
$^d$ {\em Institute for Theoretical Physics, ETH Zurich}
\\
{\em Wolfgang-Pauli-Strasse 27, 8093 Zurich, Switzerland}
\\
\\
$^e${\em Instituto de F\'isica Te\'orica, Unesp,}
\\
{\em S\~ao Paulo, 01140-070, Brazil}
}
\begin{document}
\maketitle
\begin{abstract}
We study different entropies for coherent states representing the geometry of
spherically symmetric compact systems.
We show that the thermodynamic entropy reproduces the Bekenstein-Hawking
result in the presence of thermal modes at the Hawking temperature if the
object is a black hole and saturates the Bekenstein bound for more general compact objects.
We also analyse the information entropy of the quantum coherent state without radiation
and find further support against the singular Schwarzschild geometry.
\end{abstract}
\section{Introduction}
\setcounter{equation}{0}
\label{S:intro}
Quantum aspects of gravitational collapse are among the most investigated topics in contemporary
theoretical physics.
A quantum theory of gravity is expected to eliminate the singularities predicted by general relativity,
in particular, those associated with incomplete geodesics at the final stage of the collapse of regular matter
into a black hole~\cite{HE}.
Several methods for removing the singularity in approaches to quantum gravity have been
proposed~\cite{Hajicek:2001yd,Kuntz:2019lzq,Kuntz:2017pjd,Haggard:2014rza,Kuntz:2019gup,Piechocki:2020bfo}
and the appearance of a bounce at a minimum radius is generically obtained in semiclassical
models~\cite{frolov,Casadio:1998yr,Haggard:2014rza,Schmitz:2020vdr,Casadio:2019tfz}.
These studies suggest that the collapsing matter will form a (intermediate, if not final) core of (possibly) macroscopic
size, leading to departures from the classical Schwarzschild
geometry~\cite{Calmet:2019eof,Calmet:2021stu,Calmet:2021cip,Casadio:2023uqs}.
\par
Here, we will consider the quantum realisation of Schwarzschild geometry in terms of coherent states
introduced in Ref.~\cite{Casadio:2021eio}.
Gravity in this system is described by a pure quantum state (with minimum uncertainty) for a very large
number of microscopic degrees of freedom (the virtual gravitons) depending on the
ADM~\cite{ADM} mass $M$ and the radius $R_{\rm s}$ of the compact source.
It is important to remark that there exists no state such that we can reproduce the singular
Schwarzschild metric for $R_{\rm s}\to 0$ at finite $M$. 
The matter could be described as a core of dust, like in
Refs.~\cite{Casadio:2023ymt,Casadio:2021cbv,Casadio:2022epj},
but we shall not use other information beside $M$ and $R_{\rm s}$ (see Ref.~\cite{Casadio:2023uqs}).
\par
By supplementing the coherent state with a thermal bath of Hawking quanta~\cite{Hawking:1975vcx},
one can compute the thermodynamic entropy of the system following the same procedure that was employed
in Ref.~\cite{Casadio:2015bna} for corpuscular black holes~\cite{Dvali:2011aa}.
At leading order, the result reproduces the Bekenstein-Hawking expression~\cite{bekenstein} for evaporating
black holes of large mass and saturates the Bekenstein bound~\cite{Bekenstein:1980jp} for more general
compact objects.
Moreover, sub-leading logarithmic corrections suggest that the specific heat vanishes and evaporation stops
around the Planck scale, albeit this would occur beyond the regime of validity of our approximations.
\par
The coherent Schwarzschild geometry without radiation is described by a pure quantum state
and its thermodynamic entropy would of course vanish.
Several measures of information entropy~\cite{Shannon:1948zz}
have been proposed that can be applied to pure states.
Among those, the differential configurational entropy (DCE)~\cite{Gleiser:2011di} is designed to measure
the number of bits necessary to construct a field configuration out of wave modes
in the continuum limit~\cite{Gleiser:2018jpd} and have been employed to investigate gravitational systems
and quantum field theories~\cite{Gleiser:2015aav,DaRocha:2019fjr,Gleiser:2018kbq,Bazeia:2021stz,Braga:2016wzx,Braga:2019jqg,
Lee:2017ero,Bernardini:2018uuy,Barbosa-Cendejas:2018mng,Ferreira:2019inu}.
Since the coherent Schwarzschild geometry is built in a Fock space of free (scalar) gravitons,
the DCE appears naturally suited for analysing its information content.
We shall find that the DCE further supports the non-existence of singular configurations with
$R_{\rm s}\to 0$ at finite $\Rh=2\,\gn\,M$.
Moreover, the result we will obtain here complements the analogous estimate
for the dust core in Ref.~\cite{Casadio:2022pla}.
%
%
%
%
%
%
%
%
\section{Quantum compact objects and black holes}
\setcounter{equation}{0}
\label{S:qbh}
We start here by reviewing the description of the Schwarzschild geometry in terms of coherent states
generated by a compact matter source of ADM mass $M$ and areal radius $r=R_{\rm s}$.
We then compute the thermodynamic entropy, by including a thermal spectrum of gravitons
like in Ref.~\cite{Casadio:2015bna} (see also Refs.~\cite{Casadio:2014vja,Casadio:2015lis}),
and the DCE for the background geometry alone.
We are eventually interested in computing the entropies in the black hole limit,
$R_{\rm s}\to \Rh=2\,\gn\,M$.~\footnote{We shall use units with $c=1$, the Newton constant
$\gn=\lp/\mpl$ and Planck constant $\hbar=\lp\,\mpl$, where $\lp$ is the Planck length and $\mpl$
the Planck mass.}
\subsection{Quantum Schwarzschild background}
\label{S:coherentS}
The static Schwarzschild metric,
\be
\d s^{2}
=
-\left(1-\frac{2\,\gn\,M}{r}\right)
\d t^{2}
+\left(1-\frac{2\,\gn\,M}{r}\right)^{-1}
\d r^{2}
+r^{2}
\left(\d\theta^{2}+\sin^{2}\theta\, \d\phi^{2}\right)
\ ,
\label{schw}
\ee
can be obtained from coherent states of a scalar field $\Phi=V_{\rm S}/\sqrt{\gn}$
representing virtual gravitons~\cite{Casadio:2021eio},
where
\be
V_{\rm S}
=
\frac{1}{2}\left(1-g_{tt}\right)
=
\frac{\gn\,M}{r}
\ .
\label{Vs}
\ee
We regard the vacuum $\ket{0}$ of $\Phi$ as the quantum state of a truly
empty spacetime, in which no modes of matter or gravity are excited.
It is therefore natural to quantise $\Phi$ as a massless field satisfying the free wave equation in
Minkowski spacetime~\footnote{This approach is similar to the teleparallel gravity equivalent of
General Relativity.
Moreover, including a time dimension remains formal in the absence of evolution.}
\be
\left[
-\frac{\partial^2}{\partial t^2}
+
\frac{1}{r^2}\,\frac{\partial}{\partial r}
\left(r^2\,\frac{\partial}{\partial r}\right)
\right]
\Phi(t,r)
=
0
\ ,
\label{KG}
\ee
whose normal modes can be conveniently written as
\be
u_{k}(t,r) = e^{-i\,k\,t}\,j_0(k\,r)
\ ,
\label{uj0}
\ee
where $j_0={\sin(k\,r)}/{k\,r}$ are spherical Bessel functions satisfying
\be
4\,\pi
\int_0^\infty
r^2\,\d r\,
j_0(k\,r)\,j_0(p\,r)
=
\frac{2\,\pi^2}{k^2}\,\delta(k-p)
\ .
\ee
We can now introduce the usual annihilation operators $\a_k$ and creation operators $\ac_k$
for these modes.
The quantum Minkowski vacuum is then defined by $\a_k\ket{0}=0$ and the corresponding Fock space
is built as usual.
\par
Classical configurations of the scalar field that can be realised in the quantum theory
must correspond to suitable states in this Fock space,
and a natural choice is given by coherent states
\be
\ket{g}
=
e^{-N_{\rm G}/2}\,
\exp\left\{
\int_0^\infty
\frac{k^2\,\dd k}{2\,\pi^2}\,
g_k\,
\ac_k
\right\}
\ket{0}
\label{gstate}
\ee
such that
\be
\sqrt{\frac{\lp}{\mpl}}
\bra{g}\hat{\Phi}(t,r)\ket{g}
=
V_{\rm S}(r)
=
\int_0^\infty
\frac{k^2\,\dd k}{2\,\pi^2}\,
\tilde V_{\rm S}(k)\,
j_0(k\,r)
\ .
\label{expecphi}
\ee
The latter condition determines the occupation numbers for each mode $k$ as
\be
g_k
=
\sqrt{\frac{k}{2}}\,
\frac{\tilde V_{\rm S}(k)}{\lp}
=
-\frac{4\,\pi\,M}{\sqrt{2\,k^3}\,\mpl}
\ .
\label{gkN}
\ee
It is now crucial that the state~\eqref{gstate} is well-defined only if it is normalisable, that is
if the total occupation number
\be
\Ng
=
\int_0^{\infty}
\frac{k^2\,\dd k}{2\,\pi^2}\,
g_k ^2
\label{NGN}
\ee
is finite.
However, the integral in Eq.~\eqref{NGN} with the occupation numbers~\eqref{gkN} diverges
both in the infrared (IR) and the ultraviolet (UV).
This implies that no quantum state exists in our Fock space which can reproduce $V_{\rm S}$
exactly.
Any quantum realisation of the Schwarzschild geometry must therefore contain occupation
numbers $g_k$ which differ from those in Eq.~\eqref{gkN} for $k\to 0$ and $k\to\infty$,
to make the quantum state normalisable.
The explicit form of such proper occupation numbers will depend on the state of matter
in the sourcing compact object or black hole.
Instead of assuming a particular description of such sources, we will try to derive general conclusions
from qualitative arguments.
In particular, the IR divergence occurs simply due to the assumption that the system is completely static
and the potential $V_{\rm S}$ extends to infinite distance from the source centred at $r=0$.
To cure the IR divergence we can introduce a cut-off $k_{\rm IR} = 1/ 2\,\Rinf$ to account for the
necessarily finite life-time $\tau\sim \Rinf$ of any realistic source.
The UV divergence is instead due to the behaviour of $V_{\rm S}$ for $r\to 0$
and is not present if the source is extended.
This allows us to connect the geometry with the size of the compact source
by setting $k_{\rm UV}=1/ 2\,R_{\rm s}$, which returns the correct total (thermodynamic)
energy for the background, as we shall see below.
\par
The total occupation number with the above prescriptions reads
\be
N_{\rm G}
=
4\,\frac{M^2}{\mpl^2}
\int_{k_{\rm IR}}^{k_{\rm UV}}
\frac{\dd k}{k}
=
4\,\frac{M^2}{\mpl^2}\,
\ln\left(\frac{R_\infty}{R_{\rm s}}\right)
\ ,
\label{N_G}
\ee
and we have again recovered a scaling of the mass compatible with the horizon area quantisation~\cite{bekenstein}.
Moreover, the average radial momentum is given by
\be
\expec{k}
=
4\,\frac{M^2}{\mpl^2}
\int_{k_{\rm IR}}^{k_{\rm UV}}
\dd k
=
2\,\frac{M^2}{\mpl^2}
\left(
\frac{1}{R_{\rm s}}
-
\frac{1}{R_\infty}
\right)
\ ,
\label{<k>}
\ee
and the typical wavelength $\lambda_{\rm G}={N_{\rm G}}/{\expec{k}}\sim {\lp\,M}/{\mpl}$ also reproduces
the scaling found in the corpuscular picture of black holes.
\par
We can next recompute the expectation value of the scalar field in the proper quantum state $\ket{g}$ and
find
\be
V_{\rm QS}
\simeq
\int_{k_{\rm IR}}^{k_{\rm UV}}
\frac{k^2\,\dd k}{2\,\pi^2}\,
\tilde V_{\rm S}(k)\,j_0(k\,r)
\simeq
V_{\rm S}
\left\{
1
-\left[1-
\frac{2}{\pi}\,{\rm Si}\left(\frac{r}{R_{\rm s}}\right)
\right]
\right\}
\ ,
\label{Vq}
\ee
where ${\rm Si}$ denotes the sine integral function.
We remark that $V_{\rm QS}$ displays oscillations around the expected classical behaviour $V_{\rm S}$
which become smaller and smaller for decreasing values of $R_{\rm s}$ in the region $r>\Rh$.
\subsection{Thermodinamic entropy}
\label{S:Thentropy}
Like in Ref.~\cite{Casadio:2015bna}, we start by considering a system of a large number
$N$ of scalar particles, $i=1,\ldots,N$, whose individual dynamics is determined by a Hamiltonian $H_i$.
We assume the single-particle Hilbert space contains the coherent ground state $\ket{g}$
defined previously, and a gapless continuous spectrum of energy eigenstates $\ket{\omega_i}$,
such that
\be
\hat H_i\ket{\omega_i}
=
\omega_i\ket{\omega_i}
\ .
\ee
This continuous spectrum is meant to reproduce the Hawking radiation~\cite{Hawking:1975vcx}
that will escape the coherent state and is characterised by a temperature $T=\beta^{-1}$, to wit
\be
\ket{\psi^{(i)}}
=
\mathcal{N}
\int_{\omega_{\rm c}}^\infty
\d \omega_i\,
{\frac{\omega_i-\omega_{\rm c}}{\left\{\exp\left[\beta\left(\omega_i-\omega_{\rm c}\right)\right]-1\right\}^{1/2}}
\Ket{\omega_i}}
\equiv
\int
\d\mu_i\,\ket{\omega_i}
\ ,
\label{ThermalWF}
\ee
where $\mathcal{N}=[2\,\beta^{-3}\,\zeta(3)]^{-1/2}$ is a normalisation factor, for $\zeta(3)$ being the Ap\'ery's constant, and $\omega_{\rm c}$ the minimum energy
for Hawking modes to escape the background state.
We will fix both $\beta$ and $\omega_{\rm c}$ later.
Each particle is then assumed to be in a state given by a superposition
of $\ket{g}$ and the continuous spectrum, namely
\be
\ket{\Psi^{(i)}}
=
\frac{\ket{g}+\bar\gamma\ket{\psi^{(i)}}}
{\sqrt{1+\bar\gamma^2}}
\ ,
\ee
where $0\le \bar\gamma\ll 1$ is a real parameter that weights the relative probability
amplitude for each particle to be in the thermal rather than background state.
\par
The total wave-function of the system of $N$ such bosons will correspondingly
be approximated by the totally symmetrised product
\be
\ket{\Psi}
\simeq
\frac{1}{N!}\,
\sum_{\{\sigma_i\}}^N
\left[
	\bigotimes_{i=1}^N
	\,
	\ket{\Psi^{(i)}}
	\right]
\ ,
\label{Nstate}
\ee
where $\sum_{\{\sigma_i\}}^N$ denotes the sum over all of the $N!$ permutations
$\{\sigma_i\}$ of the $N$ terms inside the square brackets.
Upon expanding in powers of $0\le\bar\gamma\ll 1$, we obtain
\be
\left(1+\bar\gamma^2\right)^{N/2}
\ket{\Psi}
\simeq
\frac{1}{N!}
\sum_{\{\sigma_i\}}^N
\left[
	\bigotimes_{i=1}^N
	\ket{g}
	\right]
+
\bar\gamma\,
\frac{\mathcal{N}}{N!}\,
\sum_{\{\sigma_i\}}^{N}
\left[
	\bigotimes_{i=2}^{N}
	\ket{g}
	\otimes
	\int
	\d \mu_1\,
	\ket{\omega_1}
	\right]
\ ,
\ee
where we omitted all the terms of order $\bar\gamma^J$ with $J=2,\ldots,N$.
\par
The spectral decomposition of this $N$-particle state can be obtained by defining the
total Hamiltonian simply as the sum of $N$ single-particle Hamiltonians,
\be
\hat H
=
\bigoplus_{i=1}^N
\hat H_i
\ .
\ee
Since we assumed $\bar\gamma\ll 1$, we again keep only terms up to first order in $\bar\gamma$,
which leads to
\be
\bra{\Psi}\hat H\ket{\Psi}
\simeq
\frac{1}{1+\gamma^2}
\left[
	\bra{g}\hat H\ket{g}
	+
	\gamma^2
	\bra{\psi}\hat H\ket{\psi}
	\right]
\ ,
\label{expE}
\ee
where $\gamma\sim\bar\gamma$~\cite{Casadio:2015bna}.
The background value is obtained from Eq.~\eqref{<k>} and reads
\be
\bra{g}\hat H\ket{g}
=
\hbar\,\expec{k}
\simeq
2\,\frac{\lp\,M^2}{R_{\rm s}\,\mpl}
\ ,
\label{gHg}
\ee
in which we took the limit $R_\infty\to \infty$.
Since the radiation is produced by the Hawking effect, it is natural to assume that the temperature
$\beta^{-1}$ is given by the surface gravity $\kappa$ of the compact object of radius $R_{\rm s}$, that is
\be
\beta^{-1}
=
\frac{\hbar\,\kappa}{2\,\pi}
\simeq
\frac{\lp^2\,M}{2\,\pi\,R_{\rm s}^2}
\ ,
\label{mR}
\ee
and that the minimum energy for Hawking quanta to escape is given by the background energy,
\be
\omega_{\rm c}
\simeq
\bra{g}\hat H\ket{g}
\simeq
\frac{\lp^3\,M^3\,\beta}{\pi\,\mpl\,R_{\rm s}^3}
\ .
\label{ORH}
\ee
In the black hole limit, $R_{\rm s}\to \Rh=2\,\gn\,M$, we thus recover the Hawking temperature
\be
\beta^{-1}
\to
\beta_{\rm H}^{-1}
=
\frac{\mpl^2}{8\,\pi\,M}
\label{mH}
\ee
and $\omega_{\rm c}\to M$.
\par
The contribution from the thermal radiation is given by
\be
\bra{\psi}\hat H\ket{\psi}
&\!\!\simeq\!\!&
\mathcal{N}^2
\int_{\omega_{\rm c}}^\infty
\frac{(\omega-\omega_{\rm c})^2}{\mathrm{exp}\left\{\beta\left(\omega-\omega_{\rm c}\right)\right\}-1}
\, \omega \, \d \omega
\nonumber
\\
&\!\!\simeq\!\!&
\omega_{\rm c}
+
\frac{\pi^4}{30\,\zeta(3)\,\beta}
\nonumber
\\
&\!\!\simeq\!\!&
\frac{\lp^3\,M^3\,\beta}{\pi\,\mpl\,R_{\rm s}^3}
+
\frac{\pi^4}{30\,\zeta(3)\,\beta}
\ ,
\ee
where we used the expression for $\omega_{\rm c}$ in Eq.~\eqref{ORH}.
The total energy is thus approximated by
\be
\bra{\Psi}\hat H\ket{\Psi}
\simeq
\frac{\lp^3\,M^3\,\beta}{\pi\,\mpl\,R_{\rm s}^3}
+
\frac{\gamma^2\,\pi^4}{30\,\zeta(3)\,\beta}
\ ,
\label{expER}
\ee
and we notice that the Eq. (\ref{expER}) in the black hole limit $R_{\rm s}\to\Rh$ yields
\be
\bra{\Psi}\hat H\ket{\Psi}
\to
\frac{\mpl^2\,\beta_{\rm H}}{8\,\pi}
+
\frac{\gamma^2\,\pi^4}{60\,\zeta(3)\,\beta_{\rm H}}
=
M
+
\frac{\gamma^2\,\pi^3\,\mpl^2}{240\,\zeta(3)\,M}
\ .
\label{Ebh}
\ee
\par
We can use Eq.~\eqref{expER} to estimate the partition function of the system
according to
\be
\expec{\hat H}
=
- \frac{\partial}{\partial \beta} \ln\left[{Z(\beta)}\right]
\ .
\ee
We then find
\be
\ln\left[{Z(\beta)}\right]
\simeq
-\frac{\lp^3\,M^3\left(\beta^2-\beta_*^2\right)}{2\,\pi\,\mpl\,R_{\rm s}^3}
-
\frac{\gamma^2\,\pi^4}{30\,\zeta(3)}\,\ln\left(\frac{\beta}{\beta_*}\right)
\ ,
\ee
where $\beta_*$ is an integration constant.
The canonical entropy is then given by
\be
S(\beta)
\simeq
\beta^2\,
\frac{\partial F}{\partial \beta}
\ ,
\ee
where $F(\beta)=-(1/\beta)\,\ln(Z)$ is the Helmoltz free energy.
It is straightforward to get
\be
S(\beta)
&\!\!\simeq\!\!&
\frac{\lp^3\,M^3\left(\beta^2+\beta_*^2\right)}{2\,\pi\,\mpl\,R_{\rm s}^3}
+
\frac{\gamma^2\,\pi^4}{30\,\zeta(3)}
\left[1-\ln\left(\frac{\beta}{\beta_*}\right)\right]
\nonumber
\\
&\!\!\simeq\!\!&
\frac{2\,\pi\,R_{\rm s}\,M}{\lp\,\mpl}
\left(
1+\frac{\lp^4\,M^2\,\beta_*^2}{4\,\pi^2\,R_{\rm s}^4}
\right)
+
\frac{\gamma^2\,\pi^4}{30\,\zeta(3)}
\left[1-\ln\left(\frac{2\,\pi\,R_{\rm s}^2}{\lp^2\,M\,\beta_*}\right)\right]
\ .
\label{SbRH}
\ee
\par
For $R_{\rm s}\to \Rh$, the above expression yields
\be
S(\beta)
\to
\frac{4\,\pi\,M^2}{\mpl^2}
\left(
1
+
\frac{\mpl^4\,\beta_*^2}{64\,\pi^2\,M^2}
\right)
+
\frac{\gamma^2\,\pi^4}{30\,\zeta(3)}
\left[1-\ln\left(\frac{8\,\pi\,M}{\mpl^2\,\beta_*}\right)\right]
\ .
\ee
On further assuming $\beta_*^{-1}\sim\mpl$, we finally obtain the entropy
\be
S(\beta_{\rm H})
&\!\!\simeq\!\!&
\frac{4\,\pi\,M^2}{\mpl^2}
\left(
1
+
\frac{\mpl^2}{M^2}
\right)
+
\frac{\gamma^2\,\pi^4}{30\,\zeta(3)}
\left[1-\ln\left(\frac{8\,\pi\,M}{\mpl}\right)\right]
\nonumber
\\
&\!\!\simeq\!\!&
\frac{\mpl^2\,\beta_{\rm H}^2}{16\,\pi}
-
\frac{\gamma^2\,\pi^4}{30\,\zeta(3)}
\ln\left(\mpl\,\beta_{\rm H}\right)
\ ,
\label{Sbh}
\ee
where we neglected terms of order $\mpl/M\ll 1$ in the last line.
\par
The leading-order term in Eq.~\eqref{Sbh} exactly reproduces the Bekenstein-Hawking
expression~\cite{bekenstein}
\be
S_{\rm BH}
=
\frac{\mathcal A}{4\,\lp^2}
=
\frac{\pi\,\Rh^2}{\lp^2}
\ ,
\label{SBH}
\ee
where $\mathcal A$ is the horizon area.
Since the total occupation number $N=N_{\rm G}$ is given in Eq.~\eqref{N_G}, we conclude
that the entropy is directly related to the normalisation of the coherent state.
Moreover, for the same choice of $\beta_*$ and to the same leading order,
the entropy~\eqref{SbRH}  reads
\be
S(\beta)
\simeq
\frac{2\,\pi\,R_{\rm s}\,M}{\lp\,\mpl}
\ ,
\ee
which is the famous Bekenstein bound~\cite{Bekenstein:1980jp}.
\par
The correction to the energy in Eq.~\eqref{Ebh} provides a positive contribution to the specific heat,
\be
C_{\rm BH}
=
-\beta_{\rm H}^2\,
\frac{\partial \expec{\hat H}}{\partial \beta_{\rm H}}
\simeq
-\frac{\mpl^2\,\beta_{\rm H}^2}{8\,\pi}
+
\frac{\gamma^2\,\pi^4}{30\,\zeta(3)}
\ ,
\ee
which makes the specific heat vanish, albeit for a mass $M\sim \gamma\,\mpl \ll \mpl$
outside the regime of approximation employed here.
This suggests that the evaporation slows down when the mass approaches the Planck scale,
like one expects from the microcanonical description of evaporating
black holes~\cite{Casadio:1997yv}. 
\subsection{Information entropy}
\label{S:Centropy}
If we omit the Hawking radiation, there is no thermal ensemble to speak of, and the background
geometry is described by the pure quantum state~\eqref{gstate}.
For pure states, we can still evaluate some information entropy~\cite{Shannon:1948zz}
based on the fact that the coherent states~\eqref{gstate} are built in the Fock space 
of Minkowski free wave modes~\eqref{uj0}.
In particular, we shall employ the DCE which precisely measures the amount of information
that is needed to assemble any field configuration from such wave modes in momentum space,
also encompassing the information complexity.
We remark that, in general, the DCE vanishes when wave modes contribute equally,
whereas a non-uniform distribution leads to increasing values~\cite{Gleiser:2018jpd}.
\par
In order to compute the DCE for a continuous wave spectrum like the one in Eq.~\eqref{uj0},
it is first convenient to introduce a dimensionless momentum variable $\tilde k=\ell\,k$,
where $\ell$ is a reference length scale.~\footnote{Note that the numerical values of the DCE
will depend on the particular choice of $\ell$, but we are here primarily interested in the 
dependence of the DCE on the macroscopic parameters of the system, that is
$M$ and $R_{\rm s}$ (and $R_\infty$).}
The DCE for a spherically symmetric system is then given by
\be 
S_{\rm{DCE}}
=
-\int
\frac{\d \tilde k}{2\,\pi^2}\,
f_{\tilde k}\, \ln\left(f_{\tilde k}\right)
\ ,
\label{eq:centropy}
\ee
where 
\begin{equation}
\label{mfraction}
f_{\tilde k}
=
\left(
\int
\frac{\d {\tilde k}}{2\,\pi^2}\,\rho_{\tilde k}^{2}
\right)^{-1}
\rho_{\tilde k}^2
\end{equation}
is the modal fraction, which characterises the contribution of distinct wave modes in momentum space.
The modal fraction encodes the way a given mode $k$ contributes to the power spectrum,
which describes fluctuations of the occupation numbers and represents the 2-point correlator
in Fourier space.
\par
Considering the occupation numbers~\eqref{gkN} and their integral~\eqref{N_G},
we define
\be
\rho_{\tilde k} = \tilde k \,g _{\tilde k}
\ .
\label{rho}
\ee
The modal fraction for our coherent state then reads
\begin{equation}
f_{\tilde k}
=
\frac{\rho_{\tilde k}^{2}}{N_{\rm G}}
=
\frac{2\,\pi^{2}}{{\tilde k}\,\ln(R_{\infty}/R_{\rm s})}
\ .
\label{bh:mfraction}
\end{equation}
The above modal fraction measures the contribution of a range of modes, between the IR and UV cut-offs
in Eqs.~\eqref{N_G}-\eqref{Vq}, to the shape of $\rho_{\tilde k}$, or equivalently, of the occupation numbers.
We then obtain 
\be
S_{\rm DCE}
&\!\!=\!\!&
-\int_{\ell/R_{\infty}}^{\ell/R_{\rm s}}
\frac{\d \tilde k}
{\tilde k\,\ln\left(R_{\infty}/R_{\rm s}\right)}\,
\ln\left[\frac{2\,\pi^{2}}{\tilde k\,\ln\left({R_{\infty}}/{R_{\rm s}}\right)}\right]
\nonumber 
\\
&\!\!=\!\!&
-\frac{1}{2}\,
\ln\left(
\frac{4\,\pi^{4}\,R_{\rm s}\,R_{\infty}}{\ell^2}
\right)
+\ln\left[\ln\left(\frac{R_{\infty}}{R_{\rm s}}\right)\right]
\ ,
\label{eq:SCE_gen}
\ee
representing the amount of information to describe the spatial profile of $\rho_{\tilde k}$ in terms of Fourier modes.
A surprising feature of the above expression is that it does not depend on the mass $M$,
unless one assumes that $\ell\sim M$, as we shall see next.
\subsubsection{Box normalisation}
\begin{figure}[H]
\begin{center}
\includegraphics[width=0.46\textwidth]{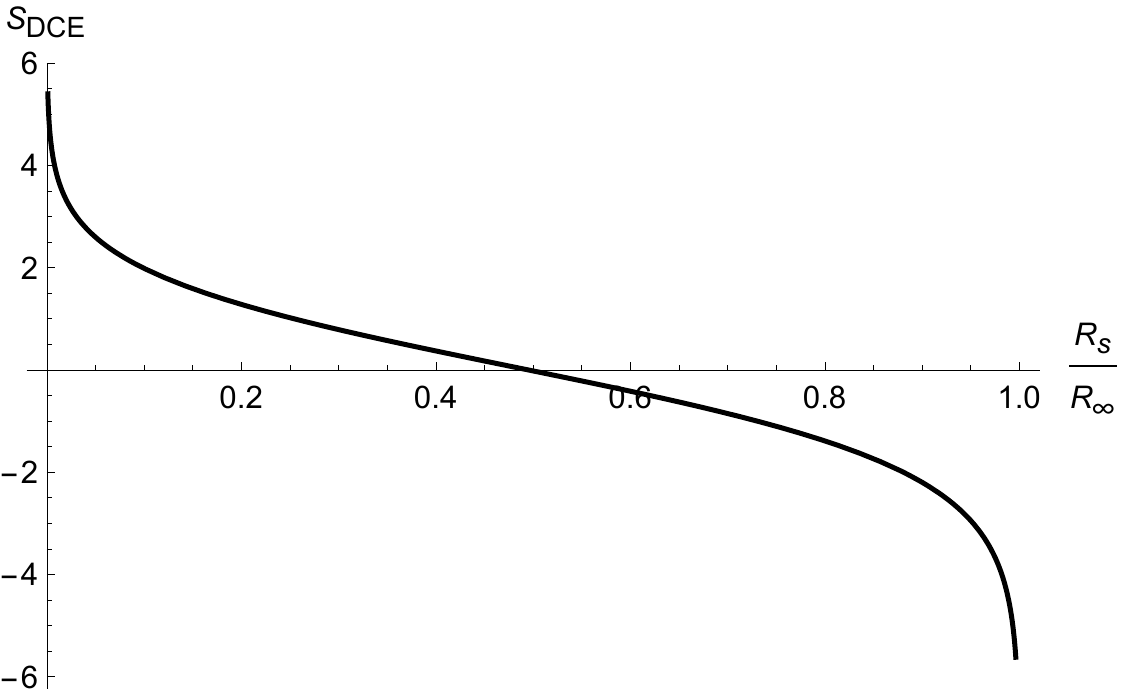}
$\ $
\includegraphics[width=0.46\textwidth]{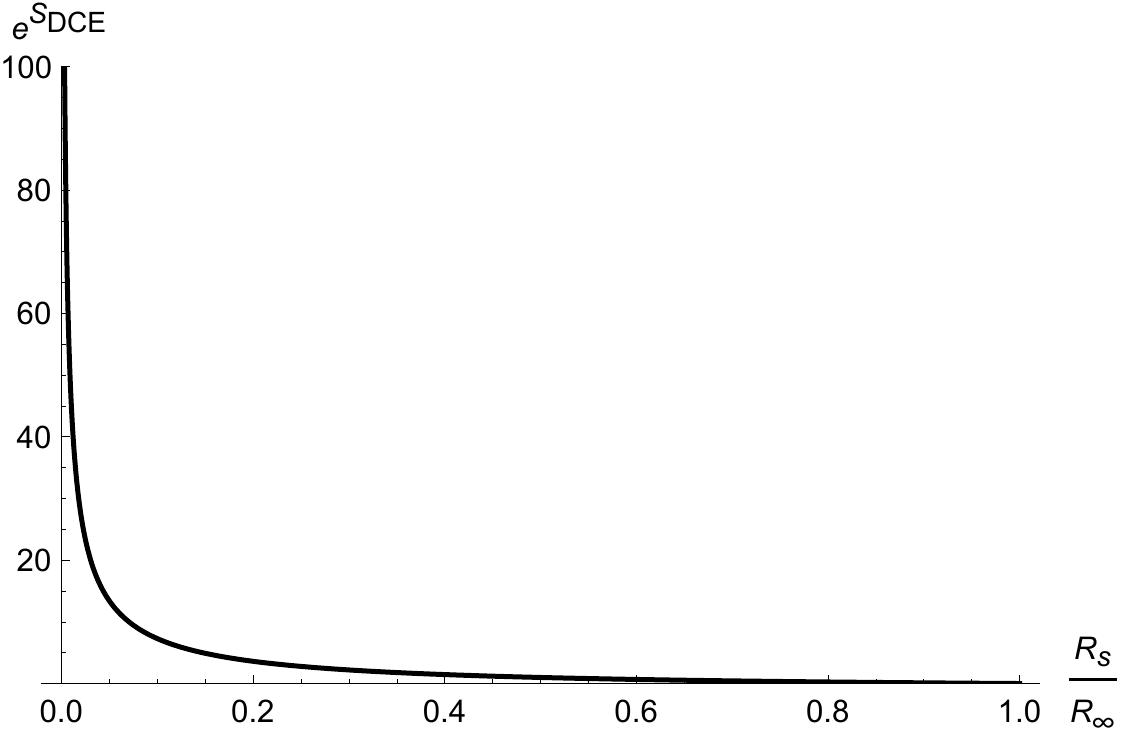}
\end{center}
\caption{Plot of the DCE in Eq.~\eqref{SceG} (left panel) and its exponential
(right panel).
The same behaviour is found in the black hole limit $R_{\rm s} \to \Rh$.}
\label{fig:Sce}
\end{figure}

The expression~\eqref{eq:SCE_gen} contains a reference length scale $\ell$.
We can first assume that $\ell$ is associated with the size $R_\infty$
of the whole system and set $\ell=4\,\pi^2\,R_\infty$.
This choice yields
\be
S_{\rm DCE}
=
\frac{1}{2}\,
\ln\left(
\frac{R_{\infty}}{R_{\rm s}}
\right)
+
\ln\left[\ln\left(\frac{R_{\infty}}{R_{\rm s}}\right)\right]
\ ,
\label{SceG}
\ee
which is plotted in the left panel of Fig.~\ref{fig:Sce}. 
This behaviour is consistent with the fact that higher values of the ratio $R_{\rm s}/\Rinf$ 
correspond to a narrower range of the wavenumber $k$ between the cut-offs $k_{\rm IR}$
and $k_{\rm UV}$.
The fewer choices are allowed for $k$, the lower the $S_{\rm DCE}$, 
corresponding to more stable configurations for the quantum coherent state.
The $S_{\rm DCE}$ diverges negatively for $R_{\rm s}/R_{\infty}=1$, a value which implies
$k_{\rm UV} = k_{\rm IR}$. 
For  $k_{\rm UV}\gg k_{\rm IR}$, corresponding to $R_{\rm s}/\Rinf \ll 1$, there is a larger range
of $k$ available to construct the quantum Schwarzschild spacetime from wave modes.
\par
In order to clarify the meaning of negative values of the DCE for $R_{\rm s}$ approaching 
$R_\infty$, we plot the exponential of the DCE in the right panel of Fig.~\ref{fig:Sce}.
This everywhere positive function decreases monotonically for increasing $R_{\rm s}$,
diverging for $R_{\rm s}\to 0$ and vanishing at $R_{\rm s}=R_{\infty}$.
If we assume that the exponential of an entropy measures the number of microscopic
configurations contributing to a given macroscopic state, this result implies that
(the gravitational field of) more compact objects contains more information
in the form of microscopic configurations. 
We also recall that the exponential of the DCE is the upper limit of the inverse squared
norm of the modal fraction~\cite{Baez}.
The right panel of Fig.~\ref{fig:Sce} therefore indicates a minimal power spectrum for
$R_{\rm s}\ll R_\infty$ and a maximal one for $R_{\rm s}\to R_\infty$.
\par
The divergence for vanishing size, in particular, would be consistent with the fact that the
coherent state~\eqref{gstate} is not well defined for $R_{\rm s}\to 0$ if $M$ is not zero.
However, in the black hole limit, $R_{\rm s} \to \Rh$, Eq.~\eqref{SceG} simply reads
\begin{equation}
S_{\rm DCE}(\Rh)
=
\frac{1}{2}\,
\ln\left(
\frac{R_{\infty}}{\Rh}
\right)
+
\ln\left[\ln\left(\frac{R_{\infty}}{\Rh}\right)\right]
\ ,
\label{SceM}
\end{equation}
which does not show any additional feature to distinguish (the gravitational field of) regular
sources from black holes.
More puzzling is the fact that the DCE in Eq.~\eqref{SceM} still diverges
for $\Rh\to 0$ (that is, $M\to 0$).
A vanishing $M$ should however correspond to Minkowski spacetime,
for which we instead expect that the DCE is zero.  
\subsubsection{Mass normalisation}
\begin{figure}[t]
\begin{center}
\includegraphics[width=0.46\textwidth]{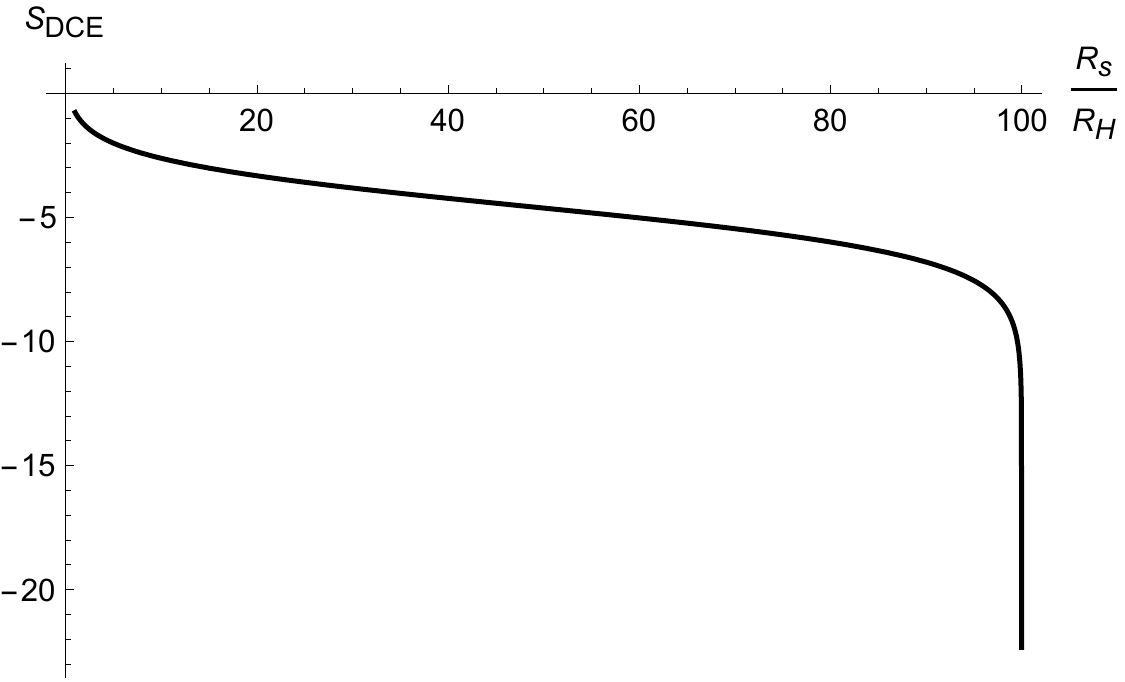}
$\ $
\includegraphics[width=0.46\textwidth]{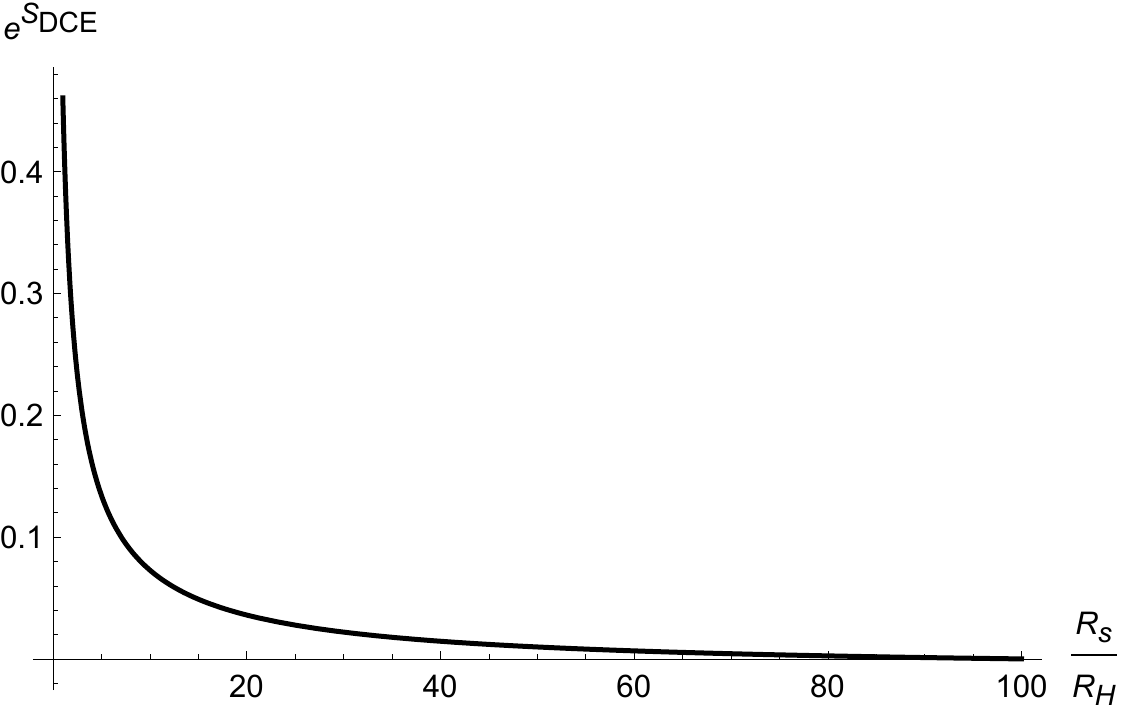}
\end{center}
\caption{Plot of the DCE in Eq.~\eqref{SceGBH} (left panel) and its exponential
(right panel) for $\Rh\le R_{\rm s}\le R_\infty=100\,\Rh$.}
\label{fig:Scebh}
\end{figure}

A DCE that knows about the mass of the compact object (or black hole) can be defined
by setting the reference scale $\ell=2\,\pi^2\,\Rh\sim M$.
In this case, we find
\be
S_{\rm DCE}
=
-\frac{1}{2}\,
\ln\left(
\frac{R_{\rm s}}{\Rh}
\right)
-
\frac{1}{2}\,
\ln\left(
\frac{R_{\infty}}{\Rh}
\right)
+
\ln\left[\ln\left(\frac{R_{\infty}}{R_{\rm s}}\right)\right]
\ .
\label{SceGBH}
\ee
An example is plotted in Fig.~\ref{fig:Scebh}, from which we see that 
the exponential of the above DCE decreases monotonically from $R_{\rm s}=\Rh$
and vanishes for $R_{\rm s}=R_\infty$. 
The left panel in Fig.~\ref{fig:Scebh} shows that the configuration $R_{\rm s}=R_{\rm H}$
is more unstable, in the context of the information entropy of the quantum coherent state,
which becomes more stable as $R_{\rm s}$ increases towards $R_\infty\sim 100\,R_{\rm H}$.
At $R_{\rm s}=R_\infty$, the quantum coherent state attains a maximal configurational stability.
\par
For $R_{\rm s}<\Rh$, Eq.~\eqref{SceGBH} behaves like the DCE in Eq.~\eqref{SceG}
(this range is not included in Fig.~\ref{fig:Scebh}).
In particular, its exponential still diverges for $R_{\rm s}/\Rh\to 0$, which is
in more explicit agreement with the result that the coherent state~\eqref{gstate} does
not exist in this limit. 
We can now argue that such a state would in fact need an infinite amount of information
to be created (by exciting modes from the Minkowski vacuum). 
The right panel of Fig.~\ref{fig:Scebh} also indicates a minimal power spectrum for
$R_{\rm s}\ll R_{\rm H}$ and a maximal one for $R_{\rm s}\to R_{\rm H}$.
\par
\begin{figure}[t]
\begin{center}
\includegraphics[width=0.46\textwidth]{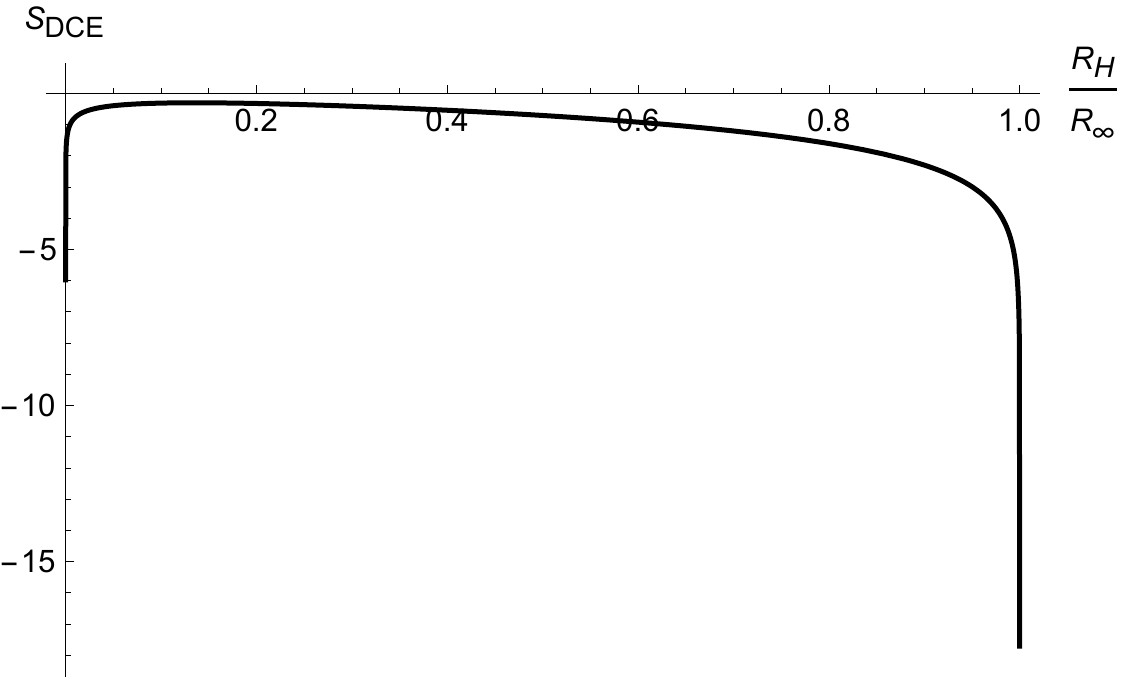}
$\ $
\includegraphics[width=0.46\textwidth]{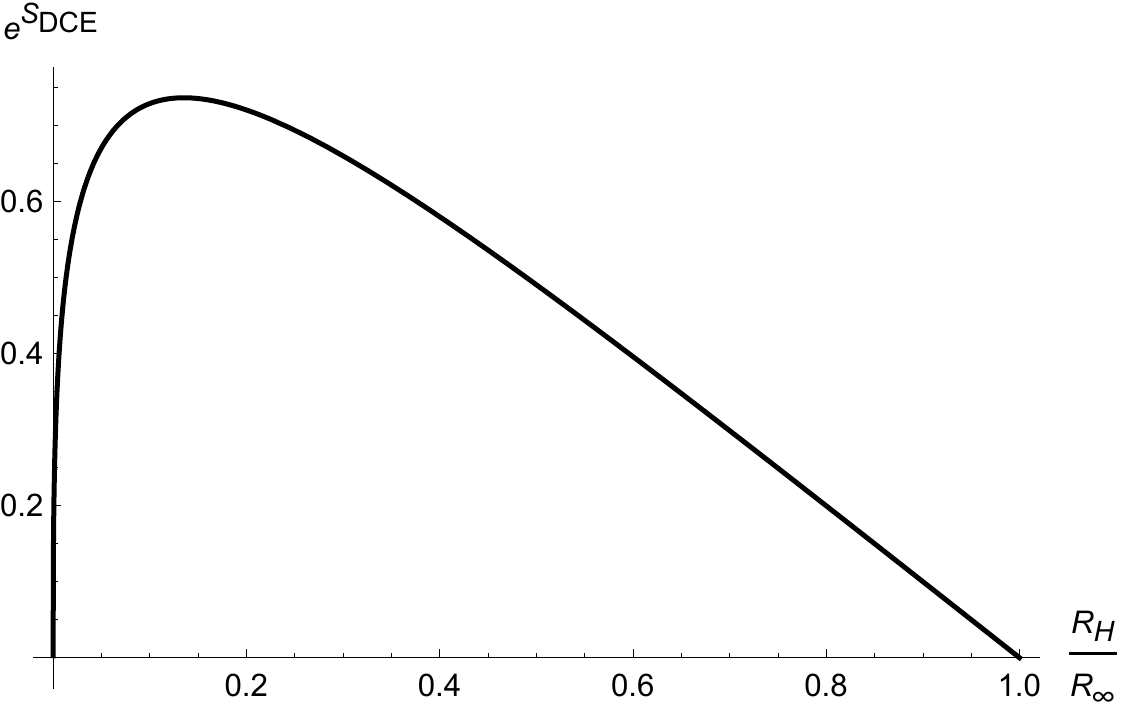}
\end{center}
\caption{Plot of the DCE in Eq.~\eqref{SceBHM} (left panel) and its exponential
(right panel).}
\label{fig:ScebhM}
\end{figure}

For $R_{\rm s}=\Rh$, Eq.~\eqref{SceGBH} simplifies to
\be
S_{\rm DCE}
=
-
\frac{1}{2}\,
\ln\left(
\frac{R_{\infty}}{\Rh}
\right)
+
\ln\left[\ln\left(\frac{R_{\infty}}{\Rh}\right)\right]
\ ,
\label{SceBHM}
\ee
which is plotted in the left panel of Fig.~\ref{fig:ScebhM} where the $S_{\rm DCE}$
is shown to diverge negatively both for $R_{\rm H}\to 0$ and $R_{\rm H}\to R_{\infty}$.
Also, for our choice of $\ell$, a maximum $S_{\rm DCE}^{\textsc{max}} = -0.307$
occurs at $R_{\rm H} = 0.135\,R_{\infty}$, indicating a point of minimum configurational
stability of the quantum coherent system.
The exponential of the DCE now vanishes both for $\Rh\to R_\infty$ and $\Rh\to 0$,
the latter case now reproducing what we would indeed expect for the vacuum Minkowski
spacetime. 
\section{Conclusions}
\label{conc}
\setcounter{equation}{0}
The Schwarzschild geometry in the quantum regime can be represented by normalisable
coherent states generated by compact matter sources, which are consistent for finite occupation
numbers. 
In order to discuss the properties of such states in some generality, we ensure that the total
occupation number~\eqref{N_G} is finite by introducing appropriate IR and UV
cut-offs~\cite{Casadio:2021eio} (and the expectation value of the scalar field, describing
virtual gravitons, in the quantum coherent states then presents oscillating corrections with
respect to the classical geometry that depend on $k_{\rm IR}$ and $k_{\rm UV}$).
\par
A total Hamiltonian for an evaporating gas of such gravitons at the Hawking
temperature~\eqref{mR} was constructed as the direct sum of $N$ single-particle Hamiltonians,
such that its expectation value on the (background) coherent state equals the ADM mass $M$
of the compact system in the black hole limit $R_{\rm s}\to\Rh$ [see Eq.~\eqref{Ebh}].
The thermodynamic canonical entropy~\eqref{SbRH} was then computed, which reproduces
the Bekenstein-Hawking entropy for black holes, and saturates the Bekenstein
bound in general, at leading order for $M\gg\mpl$.
The thermodynamic entropy is therefore directly related to the total occupation number~\eqref{N_G},
hence the normalisation of the coherent state. 
An additional (logarithmic) term is present in Eq.~\eqref{SbRH}, depending on the relative probability
amplitude for each graviton to be in the thermal state rather than the background.
This term derives from a correction to the energy that increases the specific heat
and indicates that the evaporation slows down for values of the ADM mass near the
Planck scale.
This result is corroborated by the microcanonical description of evaporating black holes.
\par
The information entropy (DCE) was used to study the quantum Schwarzschild geometry
described by pure quantum coherent states.
The DCE should measure the contribution of a range of modes, between the IR and UV cut-offs,
to the shape complexity of the occupation numbers and was analytically computed from
the modal fraction~\eqref{bh:mfraction} by introducing a reference length scale $\ell$.
The analytical expression~\eqref{eq:SCE_gen} of the DCE does not depend on the ADM mass $M$
of the quantum Schwarzschild geometry unless $\ell$ is set to be proportional to $M$.
First, a box normalisation was employed, in which the reference length $\ell$ is of the order of
the size $R_\infty\sim 1/k_{\rm IR}$ of the gravitational system.
For this case, larger values of the ratio $R_{\rm s}/R_\infty\sim k_{\rm IR}/k_{\rm UV}$,
corresponding to a narrower range of the wavenumber between the IR and the UV cut-offs,
yield lower values of the (exponential of the) DCE, which should therefore represent more stable
configurations for the quantum coherent state.
In particular, the (exponential of the) DCE diverges for $R_{\rm s}\to 0$, as does
the total occupation number~\eqref{N_G}, which signals that the coherent state becomes
highly unstable in this limit and the classical Schwarzschild geometry cannot
be realised.
For studying black holes, the choice of the mass normalisation $\ell\sim M$ appears more
interesting.
The exponential of the DCE with $R_{\rm s}=\Rh$ in Eq.~\eqref{SceBHM} shows
a maximum value for $\Rh\sim R_{\infty}/10$ and vanishes both for $\Rh\to R_\infty$ 
(when the black hole fills the entire available space) and $\Rh\to 0$ (corresponding
to a black hole with vanishing mass, hence approaching the empty Minkowski spacetime).
Such limiting configurations should therefore be the most stable, with finite mass black
holes being unstable, in agreement with the prediction of the Hawking process. 
\section*{Acknowledgments}
R.C.~is partially supported by the INFN grant FLAG and his work has also been carried out in
the framework of activities of the National Group of Mathematical Physics (GNFM, INdAM).
R.dR.~is grateful to FAPESP (Grants No.~2022/01734-7 and No.~2021/01089-1),
the National Council for Scientific and Technological Development--CNPq
(Grant No.~303390/2019-0), and the Coordination for the Improvement of Higher Education
Personnel (CAPES-PrInt~88887.897177/2023-00), for partial financial support.
P.M.~thanks FAPESP (Grants No.~2022/12401-9 and No.~2019/21281-4) for the financial support.
\appendix
\bibliographystyle{unsrt}
\end{document}